\documentclass[%
reprint,
superscriptaddress,
bibnotes,
amsmath,amssymb,
aps,
floatfix,
]{revtex4-1}
\usepackage{graphicx}
\usepackage{dcolumn}
\usepackage{bm}
\usepackage{hyperref}
\hypersetup{colorlinks=true,linkcolor=blue,citecolor=red,urlcolor=blue}
\usepackage{natbib}
\usepackage{mathrsfs,mathtools,color,wasysym,dsfont,here}
\usepackage{amsthm}
\usepackage{physics}
\usepackage[all]{xy}
\usepackage{tikz}
\usepackage{amsfonts}
\usetikzlibrary{calc}
\usetikzlibrary{arrows}
\usetikzlibrary{decorations.markings,decorations.pathmorphing}

\begin{document}

\preprint{}

\title{Onsager's Scars in Disordered Spin Chains}

\author{Naoyuki Shibata}
\email{shibata-naoyuki@g.ecc.u-tokyo.ac.jp}
\affiliation{Department of Physics, Graduate School of Science, The University of Tokyo, 7-3-1 Hongo, Tokyo 113-0033, Japan
}
\author{Nobuyuki Yoshioka}
\affiliation{Department of Physics, Graduate School of Science, The University of Tokyo, 7-3-1 Hongo, Tokyo 113-0033, Japan
}
\author{Hosho Katsura}
\affiliation{Department of Physics, Graduate School of Science, The University of Tokyo, 7-3-1 Hongo, Tokyo 113-0033, Japan
}
\affiliation{Institute for Physics of Intelligence, The University of Tokyo, 7-3-1 Hongo, Tokyo 113-0033, Japan
}
\affiliation{Trans-scale Quantum Science Institute, The University of Tokyo, 7-3-1 Hongo, Tokyo 113-0033, Japan}

\date{\today}

\begin{abstract}
    We propose a class of non-integrable quantum spin chains that exhibit quantum many-body scars even in the presence of disorder. With the use of the so-called Onsager symmetry, we construct scarred models for arbitrary spin quantum number $ S $. There are two types of scar states, namely, coherent states associated with an Onsager-algebra element and one-magnon scar states. While both of them are highly-excited states, they have area-law entanglement and can be written as a matrix product state. Therefore, they explicitly violate the eigenstate thermalization hypothesis. We also investigate the dynamics of the fidelity and entanglement entropy for several initial states. The results clearly show that the scar states are trapped in a perfectly periodic orbit in the Hilbert subspace and never thermalize, whereas other generic states do rapidly. To our knowledge, our model is the first explicit example of disordered quantum many-body scarred models.
\end{abstract}

\maketitle

\textit{Introduction.---}%
	The origin of thermalization in isolated quantum systems and the role of ergodicity have been studied for a long time~\cite{Polkovnikov2011, Nandkishore2015}. Recent experimental progress in quantum engineering including ultracold atoms~\cite{Trotzky2012}, superconducting qubits~\cite{Neill2016}, trapped ions~\cite{Smith2016}, and Rydberg atoms~\cite{Bernien2017} has provided direct access to such phenomena. As a theoretical approach, several studies revealed a plausible scenario of the thermalization of quantum systems, namely, the eigenstate thermalization hypothesis (ETH). A strong form of the ETH states that all energy eigenstates are locally indistinguishable from the microcanonical ensemble~\cite{Gogolin2016, Mori2018}. Although there is no rigorous proof, it is widely believed to hold for a large class of interacting systems, as evidenced by several numerical studies~\cite{Rigol2008, Kim2014, DAlessio2016, Garrison2018}. On the other hand, a weak version of ETH, which states \textit{almost} all energy eigenstates are locally indistinguishable from the microcanonical ensemble~\cite{Biroli2010}, was proved for generic translationally invariant short-range interacting systems~\cite{Iyoda2017}. Remarkable exceptions are integrable and many-body localized (MBL) systems~\cite{Pal2010,Alba2015,Nandkishore2015,Calabrese2016}. In such systems, the existence of an extensive number of conserved quantities/integrals of motion \textit{strongly} breaks ergodicity, and therefore the weak ETH as well~\footnote{Although a generalized ETH is known to hold for generalized Gibbs ensemble (GGE)~\cite{Cassidy2011,Ishii2019}, here we mean the standard ETH in which we do not consider additional local constraints.}.

    Recently, there has been increasing interest in systems which \textit{weakly} violate ergodicity; almost all typical states thermalize rapidly, as expected in generic interacting systems, but certain special states do not or exhibit anomalously slow thermalization, which means that they obey the weak ETH but violate the strong ETH. From another perspective, most energy eigenstates have volume-law entanglement entropy (EE), whereas those special states have sub-volume-law EE. These unusual states are called quantum many-body scars (QMBS) ~\cite{Bernien2017, Turner2018, Turner2018a, James2019}. The initial experimental observation of QMBS~\cite{Bernien2017} has stimulated further theoretical studies. In particular, an effective model of this experimental setup, dubbed the PXP model~\cite{Turner2018, Turner2018a, Ho2019, Khemani2019, Lin2019}, has been intensively studied to elucidate the peculiar absence of thermalization. Another approach is to construct models with \textit{perfect QMBS}, whose exact expression can be written down and perfect revivals in the many-body quantum dynamics can be shown analytically. Some previous work revealed a situation in which scar states live in a large global angular momentum sector protected from thermalization~\cite{Shiraishi2017, Choi2019, Schecter2019}. Others studied scar states in the Affleck-Kennedy-Lieb-Tasaki (AKLT) model~\cite{Moudgalya2018, Moudgalya2018a} or constructed such AKLT-like matrix product state (MPS) scar states~\cite{Chattopadhyay2019, Iadecola2019}. Moreover, the Floquet analog of ETH violation and QMBS has also been discussed~\cite{Sugiura2019, Mukherjee2019}. Despite such intensive studies on QMBS, its general framework and origin remain unclear~\cite{Alhambra2019}. In order to gain a better understanding, analytically tractable QMBS models are much appreciated.
    
    In this Letter, we propose a new class of spin models with QMBS. The key to the construction is the so-called Onsager algebra~\cite{Onsager1944}, which originally appeared in obtaining the exact solution of a two-dimensional classical Ising model. Focusing on a certain Onsager-algebra element, we can explicitly write down a one-parameter family of scar states as an MPS with a finite bond dimension, which means scar states have area-law EE. Our model has three remarkable features: (1) the scar state in our model is not a product state such as those discussed in Ref.~\cite{Schecter2019}, but does have a finite area-law entanglement. (2) Although here we demonstrate mainly the spin quantum number $ S=1/2 $ case, $ S $ can be an arbitrary half-integer. Scar states can also be generalized to multi-parameter ones, while we explain one-parameter ones in the main text for simplicity. See Supplemental Material \footnote{See Supplemental Material for more detailed discussion of the model and its generalizations.} for these generalizations.
    (3) We do not impose translational invariance on our model. To the best of our knowledge, this is the first explicitly constructed example of the disordered QMBS model~\footnote{Although Ref.~\cite{Shiraishi2017} can be easily generalized to cases without translational invariance, it is different from our construction.}. 
	
\medskip

\textit{Onsager symmetry in spin chain.---}%
    Before defining our model that exhibits QMBS, we first introduce the following integrable Hamiltonian of the self-dual $ \mathrm{U}(1) $-invariant clock model~\cite{Vernier2019} under the periodic boundary condition
	\begin{align}
	\begin{split}
	  	\hspace{-3em}H_{n}&=-\sum_{j=1}^{L}\sum_{a=1}^{n-1}\dfrac{1}{2\sin(\pi a/n)}\left[n(-1)^a (S_j^- S_{j+1}^+)^a+\mathrm{h.c.}\right.\\
	  	&\hspace{11em}\left.+(n-2a)\omega^{a/2}\tau_j^a\right].
	\end{split}\label{eq:SD_U1i_c}
	\end{align}
	Here, $ L $ is the number of sites and assumed to be even, $ \omega=\mathrm{e}^{2\pi\mathrm{i}/n} $, and $ n $ is a dimension of each local Hilbert space $ \mathcal{H}_j\simeq \mathbb{C}^n $, and hence the total Hilbert space is $ \bigotimes_{j=1}^L \mathcal{H}_j $. The operators $ \tau_j $ and $ S_j^\pm $ act on $ \mathcal{H}_j $ as
	\begin{gather}
	    \tau=
		\begin{pmatrix}
			1&&&\\
			&\omega&&\\
			&&\ddots&\\
			&&&\omega^{n-1}
		\end{pmatrix}
		,\;
		S^+=
		\begin{pmatrix}
			0&1&&\\
			&\ddots&\ddots&\\
			&&0&1\\
			0&&&0
		\end{pmatrix},
	\end{gather}
	and $ S^-=(S^+)^\dagger $. The simplest $ n=2 $ case reduces to the $ S=1/2 $ XX model
	\begin{align}
	    H_2=\sum_{j=1}^L S_j^+ S_{j+1}^- + S_j^- S_{j+1}^+,
	\end{align}
    and the $ n=3 $ case is known as a particular case of the Fateev-Zamolodchikov model~\cite{Zamolodchikov1980, Bytsko2003}. It can be easily seen that $ H_n $ commutes with the $ \mathrm{U}(1) $-charge $ Q $:
	\begin{align}
		Q\coloneqq \sum_{j=1}^{L}S_j^z,\quad S^z=\operatorname{diag}\qty(\dfrac{n-1}{2},\dfrac{n-3}{2},\dots,-\dfrac{n-1}{2}),
	\end{align}
	which follows from $ [Q,S_j^{\pm}]=\pm S_j^{\pm} $. Note that $ S^\pm $ are not standard spin raising/lowering operators and do not obey the $ \mathrm{SU}(2) $ commutation relation, i.e.,  $ [S^+, S^-] \not\propto S^z $ (except for an $ n = 2 $ or $ 3 $ case), and the model does not have $ \mathrm{SU}(2) $ symmetry.
	
	A remarkable observation in Ref.~\cite{Vernier2019} is that $ Q $ and $ \hat{Q} $, the dual of $ Q $ obtained by the dual transformation on $ \tau $ and $ (\sigma)_{ij}=\delta_{i,j+1\mod{n}} $~\cite{Note2},
	do not commute, but generate the \textit{Onsager algebra}~\cite{Onsager1944}. One of such Onsager-algebra elements is
	\begin{align}
	    Q^+=\sum_{j=1}^{L}\sum_{a=1}^{n-1}\dfrac{(-1)^{(n+1)j+a}}{\sin(\pi a/n)}(S_j^+)^a (S_{j+1}^+)^{n-a},
	\end{align}
	which plays an important role in generating QMBS states below~\footnote{In the case of $ n=2 $, it reduces to $ Q^+=\sum_j (-1)^{j+1}S_j^+S_{j+1}^+ $. Actually, $Q^+\ket{\Downarrow} $ is known as a singular state in the context of the Heisenberg chain~\cite{Avdeev1986,Essler1992,Noh2000,Nepomechie2013}. Moreover, it is equal to the raising operator for virtual pseudo-spins in Ref.~\cite{Chattopadhyay2019}}. Because of the self-duality, $ H_n $ also commutes with $ \hat{Q} $, and therefore, all Onsager-algebra elements including $ Q^+$. Actually, the boundary condition employed here differs from Ref.~\cite{Vernier2019}, but any important commutation relations still hold with straightforward modifications~\cite{Note2}.
	
	We denote by $ \ket{p} $ ($ p=0,1,\dots,n-1 $) the eigenstate of $ S^z $ with eigenvalue $ p-(n-1)/2 $. The ferromagnetic state $ \ket{\Downarrow}\coloneqq\otimes_{j=1}^L\ket{0} $ is the eigenstate of $ H_n $ with eigenvalue $ -L\sum_{a=1}^{n-1}\frac{(n-2a)\omega^{a/2}}{2\sin(\pi a/n)} $. Since $ [Q^+, H_n]=0 $, $ (Q^+)^k\ket{\Downarrow} $ ($ k=0,\dots, \lfloor\frac{n-1}{n}L\rfloor $) are also eigenstates of $ H_n $ with the same eigenvalue. 
	
\medskip

\textit{Model and perfect scars.---}%
    Let us consider the Hamiltonian
    \begin{align}
        H_\mathrm{S}=H_n+H_{\mathrm{pert}, n}+h\sum_{j=1}^L S_j^z.
    \end{align}
    We choose $ H_{\mathrm{pert}, n} $ so as to destroy the integrability of the Hamiltonian but keep $ (Q^+)^k\ket{\Downarrow} $ to be eigenstates as follows. We introduce an (unnormalized) \textit{coherent state}
	\begin{align}
		\ket{\psi(\beta)}\coloneqq \exp(\beta^n Q^+)\ket{\Downarrow},
	\end{align}
	which is exactly written as an MPS~\cite{Note2}:
	\begin{align}
		\ket{\psi(\beta)}=\sum_{p_1,\dots,p_L}\tr(A_{p_1}B_{p_2}\dots A_{p_{L-1}} B_{p_L})\ket{p_1\dots p_L},
	\end{align}
	where $ A_p $ and $ B_p $ are $ n\times n $ matrices whose matrix elements are (using 0-based indexing)
	\begin{align}
		(A_p)_{ij}&=\beta^{p}\delta_{i,p}\delta_{j,0} +\dfrac{(-1)^{j+1}\beta^p}{\sin[\pi(n-j)/n]}\delta_{n-p, j-i},\\
		(B_p)_{ij}&=\beta^{p}\delta_{i,p}\delta_{j,0} +\dfrac{(-1)^{n-j}\beta^p}{\sin[\pi(n-j)/n]}\delta_{n-p, j-i}
	\end{align}
	for $ 0\le i,j\le n-1 $. This MPS representation reveals that particular spin configurations over three consecutive sites never appear in $ \ket{\psi(\beta)} $. In the case of $ n=2 $, for example, it is easily verified that
    \begin{align}
        \mathsf{ABA}=
        \begin{pmatrix}
            \ket{000}-\beta^2(\ket{011}-\ket{110})&\beta\ket{001}+\beta^3\ket{111}\\
            \beta\ket{100}-\beta^3\ket{111}&\beta^2\ket{101}
        \end{pmatrix},\label{eq:MPS_ABA}
    \end{align}
    where we introduce the notation
    \begin{align}
        (\mathsf{A})_{ij}=\sum_{p=0}^{n-1}(A_p)_{ij}\ket{p}, \quad (\mathsf{B})_{ij}=\sum_{p=0}^{n-1}(B_p)_{ij}\ket{p}.
    \end{align}
    One can see that any matrix elements of Eq.~(\ref{eq:MPS_ABA}) are orthogonal to both $ \ket{010} $ and $ (\ket{011}+\ket{110})/\sqrt{2} $. The same conclusion follows from $ \mathsf{BAB} $ configuration. Therefore, we consider the following perturbation up to three-body interactions:
    \begin{align}
    \begin{split}
        H_{\mathrm{pert},2}&=\sum_{j=1}^{L}\left\{c_j^{(1)}\ket{010}\bra{010}\right.\\
        &\hspace{2em}+\dfrac{c_j^{(2)}}{2}(\ket{011}+\ket{110})(\bra{011}+\bra{110})\\
        &\hspace{2em}\left. +c_j^{(3)}\qty[\ket{010}(\bra{011}+\bra{110})+\mathrm{h.c.}]\right\}_{j-1,j,j+1}.
    \end{split}\label{eq:perturbation_n2}
    \end{align}
     Note that when $ c_j^{(3)}\ne 0 $, $ H_\mathrm{S} $ does not have $ \mathrm{U}(1)$ symmetry. 
     
     Several remarks are in order. First, we emphasize that the translational invariance is not assumed for $ H_{\mathrm{S}}$. To the best of our knowledge, such models have not been explicitly constructed before this work. Second, here we introduced a one-parameter coherent state, but a parallel discussion allows us to generalize to a multi-parameter coherent state using higher Onsager algebra elements~\cite{Note2}. Third, perturbation terms for higher spin cases $ n\ge 3 $ are also obtained in a similar way~\cite{Note2}. 
    
	
	Although $ H_n $ is integrable, it is likely that the perturbation makes $ H_\mathrm{S} $ non-integrable for generic $ c_j^{(i)} $. To confirm the non-integrability of the model, we compute the level-spacing statistics of $ H_\mathrm{S} $ by exact diagonalization in the particular case where $ c_j^{(1)} $ are chosen randomly from $ [-1, 1] $ and $ c_j^{(2)}=c_j^{(3)}=0 $. Let $ E_1\le E_2\le\cdots \le E_i\le\cdots $ be eigenvalues of $ H_\mathrm{S} $ in ascending order and $ \Delta E_i=E_{i+1} - E_i $. It is well known~\cite{Casati1985, Prosen1993, Rabson2004} that $ s_i=\Delta E_i/\left< \Delta E_i \right> $ obeys the Poisson (Wigner-Dyson) distribution if $ H_\mathrm{S} $ is integrable (non-integrable), where $ \left< \Delta E_i \right> $ is an average of $ \Delta E_i $'s. The level-spacing ratio (as known as $ r $-value)~\cite{Pal2010} $ \left< r \right>=\left< \min(\Delta E_i, \Delta E_{i+1})/\max(\Delta E_i, \Delta E_{i+1}) \right> $ is often used for quantitative detection of distribution statistics; $ \left< r \right>\simeq 0.39 $ for the Poisson distribution, and $ \left< r \right>\simeq 0.53 $ for the Wigner-Dyson distribution.
	\begin{figure}
		\centering
		\includegraphics[width=1.0\linewidth]{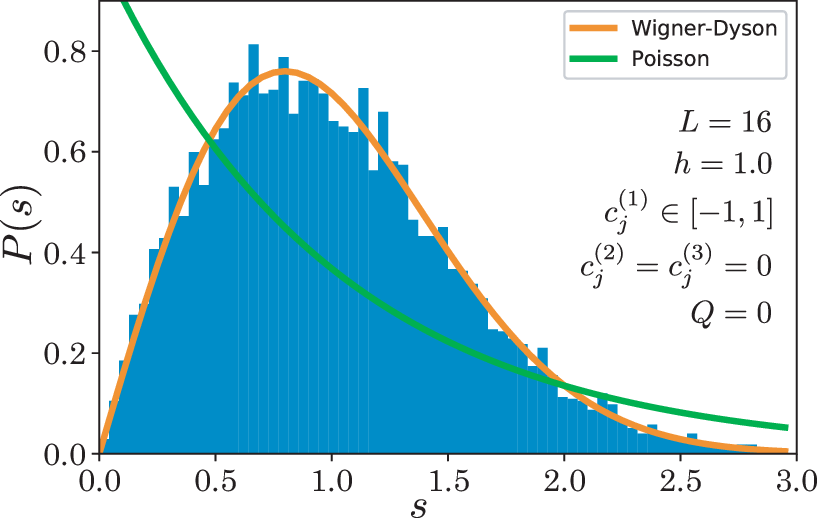}
		\caption{Level-spacing statistics in the middle half of the spectrum of $ H_\mathrm{S} $ in the $ n=2$ case. The parameters and the symmetry sector used are indicated in the figure. Each $ c_j $ is randomly chosen from $ [-1, 1] $. The probability density function of the Wigner-Dyson distribution $ P(s)=(\pi/2)s\mathrm{e}^{-\pi s^2/4} $ and the Poisson distribution $ P(s) = \mathrm{e}^{-s} $ are shown for comparison. The result agrees well with the Wigner-Dyson distribution.}
		\label{fig:level-spacing_statistics_L16_random_fig}
	\end{figure}
	The results shown in Fig.~\ref{fig:level-spacing_statistics_L16_random_fig} agree well with the Wigner-Dyson distribution, which implies the non-integrability of $H_\mathrm{S} $. Its $ r $-value $ \left< r \right>\simeq 0.5328\dots $ is also close enough to that of the Wigner-Dyson distribution~\footnote{We have investigated other different distributions of $c_j^{(i)} $ such as the normal distribution, and confirmed from the level-spacing statistics that the Hamiltonians for different distributions remain non-integrable as long as $ c_j^{(i)}=\order{1} $}.
	
	However, $ (Q^+)^k\ket{\Downarrow} $ violate the strong ETH, as they have a sub-volume-law EE even though they are excited states. In fact, one can show that an upper bound for the EE of $ (Q^+)^k\ket{\Downarrow} $ scales as $ \order{\ln L} $~\cite{Note2}. In particular, a coherent state has an area-law EE since it can be written as an MPS with finite bond dimension.
	
    \begin{figure}
        \centering
        \includegraphics[width=1.0\linewidth]{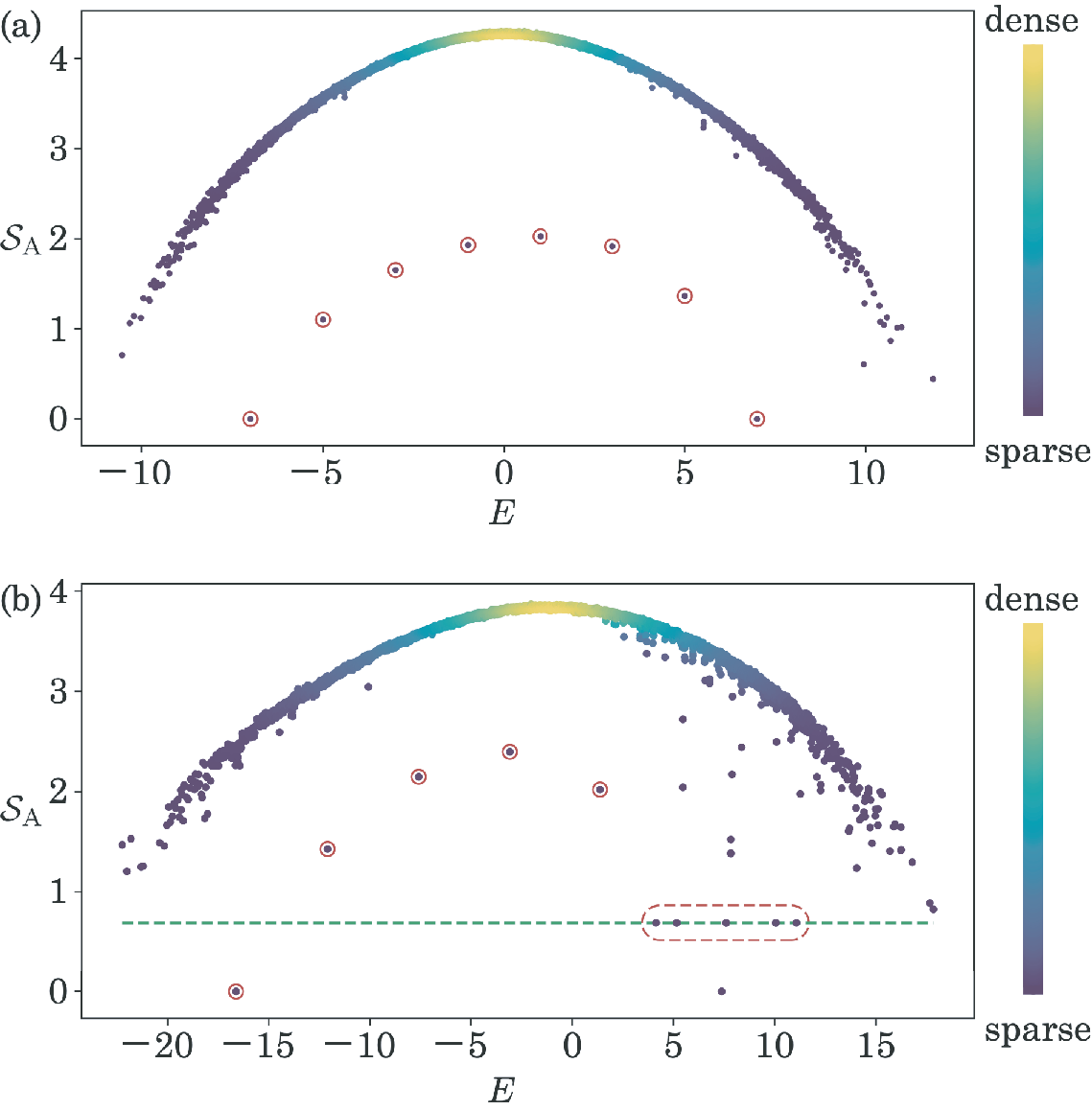}
        \caption{Half-chain bipartite EE as a function of energy $ E $ for (a) $ n=2,\; L=14,\; h=1.0 $ and (b) $n=3,\; L=8,\; h=1.5 $. Color scale for each dot indicates the density of data points. $ (Q^+)^k\ket{\Downarrow} $ are marked by red solid circles. (a) Perturbation parameters are chosen randomly as $ c_j^{(1)},c_j^{(2)},c_j^{(3)} \in [-1,1] $. (b) Perturbations are chosen not to destroy one-magnon scars indicated by the red dashed circle~\cite{Note2}. A green dashed line indicates $ \ln 2 $.}
        \label{fig:EE_n2}
    \end{figure}
		
\medskip

\textit{Entanglement entropy of our model.---}%
    The von Neumann EE is one of the measures of quantum entanglement. With respect to a bipartition of the system into subsystems A and B, the von Neumann EE of $ \ket{\phi} $ for A is defined as
    \begin{align}
        \mathcal{S}_\mathrm{A}=-\tr_\mathrm{A}(\rho_\mathrm{A} \ln \rho_\mathrm{A}),
    \end{align}
    where $ \rho_\mathrm{A}=\tr_\mathrm{B}(\ket{\phi}\bra{\phi})$ is the reduced density matrix of region A. In the following, we focus on the half-chain bipartite von Neumann EE and take the left half of the chain to be region A.

    The strong ETH states that all energy eigenstates are thermal, which implies that these energy eigenstates have volume-law entanglement~\cite{Mori2018}. Figure \ref{fig:EE_n2} shows half-chain bipartite EE for every energy eigenstate as a function of energy for (a) $ n=2 $ and (b) $ n=3 $. In both plots, a general feature of QMBS can be seen: the states in the bulk of energy spectrum have large volume-law EE, whereas some atypical states have anomalously small sub-volume-law EE, including $ (Q^+)^k\ket{\Downarrow} $ marked by red circles~\footnote{The presence of excited eigenstates in integrable models that have area-law instead of volume-law entanglement has been pointed out early in Ref.~\cite{Alba2009}.}. In (b), one can see other low EE states besides $ (Q^+)^k\ket{\Downarrow} $. In particular, EE of several states is exactly $ \ln 2 $. We identify these states as one-magnon states lying on the Hilbert subspace spanned by $ \{\ket{2\dots 212\dots 2 }\} $. Note that here $ H_\mathrm{S} $ does not have $ \mathrm{U}(1) $ symmetry. These one-magnon scars, however, disappear by adding other perturbation terms~\cite{Note2}.
\medskip

\textit{Dynamics.---}%
	The dynamics is also studied to illustrate the feature of the QMBS more explicitly. First, let us consider the dynamics of the coherent state. For the initial coherent state $ \ket{\psi_{t=0}(\beta)}=\ket{\psi(\beta)} $, it is obvious from the construction of $ H_\mathrm{S} $ that
	\begin{align}
	    \ket{\psi_t(\beta)}&=\mathrm{e}^{-\mathrm{i}H_\mathrm{S}t}\ket{\psi(\beta)}\propto\mathrm{e}^{-\mathrm{i}h\sum_j S_j^z t}\sum_{k=0}^\infty (\beta^n Q^+)^k\ket{\Downarrow}\notag\\
	    &\propto \ket{\psi(\mathrm{e}^{-\mathrm{i}ht}\beta)}.\label{eq:coherent_state_dynamics}
	\end{align}
	Although the coherent state does evolve, it returns to itself with period $ T=2\pi/(nh) $, since $ \ket{\psi(\mathrm{e}^{2\pi \mathrm{i}/n}\beta)}=\ket{\psi(\beta)} $. We emphasize that this revival is perfect, and thus the coherent state never thermalizes.
	
	\begin{figure}
	    \centering
	    \includegraphics[width=1.0\linewidth]{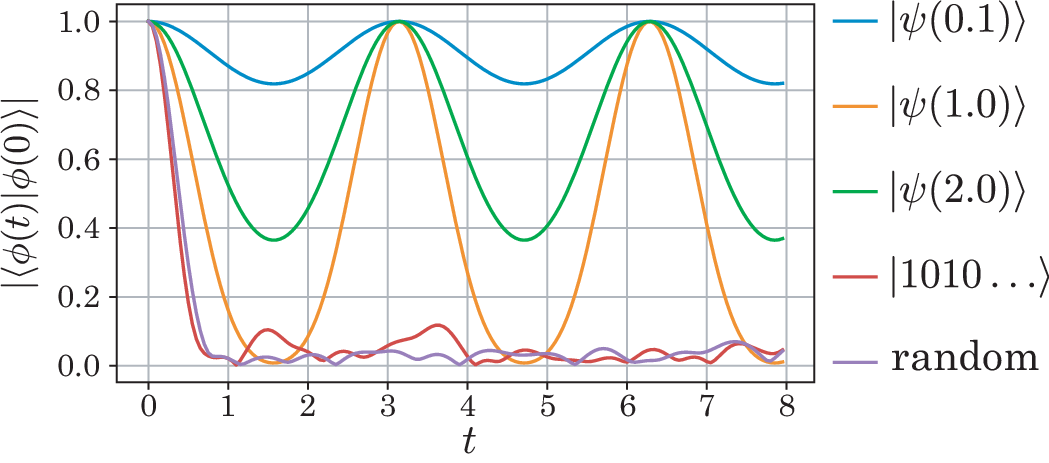}
	    \caption{Fidelity dynamics with $ n=2,\; L=10,\; h=1.0 $, and $ c_j^{(1)},c_j^{(2)},c_j^{(3)} $ chosen randomly from $ [-1,1] $. Perfectly periodic revivals can be seen in the case where the initial state is a coherent state, whereas for other typical states the fidelity decreases very quickly to $ 0 $.}
	    \label{fig:fidelity_L10_fig}
	\end{figure}
	
	\begin{figure}
	    \centering
	    \includegraphics[width=1.0\linewidth]{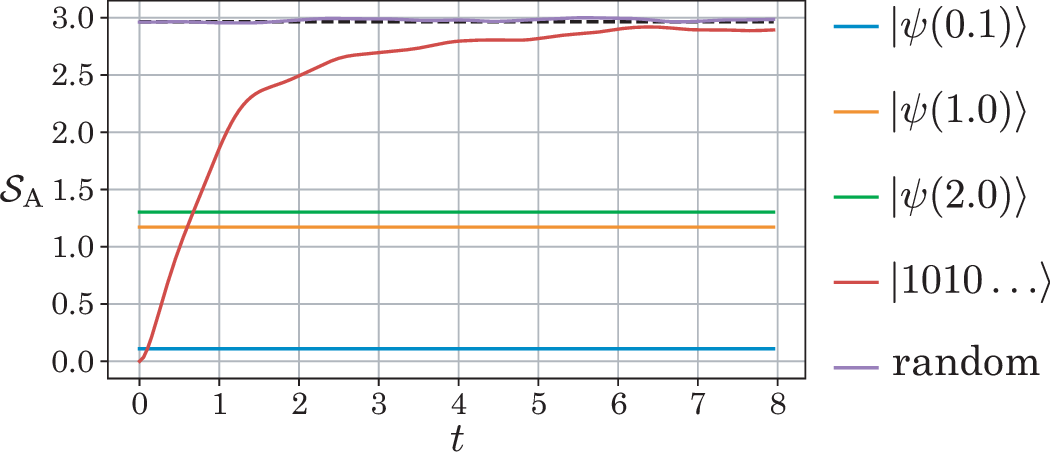}
	    \caption{Dynamics of the half-chain bipartite EE with the same setup as Fig.~\ref{fig:fidelity_L10_fig}. Initial coherent states have constant EE, while that of $ \ket{1010\dots}$ grows rapidly and saturates near the Page value denoted by the black dashed line. The EE of the random initial state almost remains at the Page value from first to last.}
	    \label{fig:EE_growth}
	\end{figure}
	
	We show in Fig.~\ref{fig:fidelity_L10_fig} the numerical results of the fidelity dynamics with several initial states $ \ket{\phi} $ defined by
	\begin{align}
	    F(t)=\abs{\bra{\phi(t)}\ket{\phi(0)}}=\abs{\bra{\phi}\mathrm{e}^{\mathrm{i}H_\mathrm{S}t}\ket{\phi}}.
	\end{align}
	When the initial states are coherent states, we can see perfectly periodic revivals of their fidelity. However, if the system starts from other generic states, its fidelity decreases rapidly to $ 0 $.
	
	We also calculate the dynamics of the half-chain bipartite EE shown in Fig.~\ref{fig:EE_growth} with the same setup as Fig.~\ref{fig:fidelity_L10_fig}. It is easy to see that the coherent state does not gain entanglement, since $ H_\mathrm{S} $ acts on $ \ket{\psi(\beta)} $ as if it is just an external field, i.e., a non-interacting term (see Eq.~(\ref{eq:coherent_state_dynamics})). On the other hand, EE of the initial product state $ \ket{1010\dots} $ grows soon and saturates near the Page value~\cite{Page1993} of a random state
	\begin{align}
	     \mathcal{S}_\mathrm{Page}=\dfrac{L}{2}\ln 2 - \dfrac{1}{2}.
	\end{align}
	From these numerical results on dynamics of the fidelity and EE, we confirm that typical states thermalize rapidly, while scar states never thermalize and violate ergodicity.
	
\medskip

\textit{Summary and outlook.---}%
    We have constructed a disordered spin chain model with QMBS with the help of the Onsager algebra. There are two types of scar states, namely, coherent scar states associated with an Onsager-algebra element and one-magnon scars. A coherent state has been written explicitly as an MPS, which implies that it has a finite but area-law EE. We have shown analytically that the coherent state undergoes a perfect revivals, and therefore never thermalizes. On the other hand, most of other generic states thermalize rapidly, as evidenced by the EE spectrum and dynamics. Although we have demonstrated our model mainly in the case of $ S=1/2 $, the results are also valid for general $ S $.
    
    Before finishing our discussion, several remarks are in order. First, Onsager scar states $ (Q^+)^k\ket{\Downarrow} $ can be prepared in a Markovian open quantum system. By taking jump operators that annihilate the coherent state, the decoherence-free subspace for this Lindblad dynamics is spanned by $ (Q^+)^k\ket{\Downarrow} $. Thus, these Onsager scar states are steady states and can be obtained through the dynamics with arbitrary initial states. Second, for the $ S=1/2 $ case, our coherent state and the ground state of the quantum lattice gas model studied in Ref.~\cite{Lesanovsky2011} are closely related to each other. In our coherent state, let us define bond variables for each bond between site $ j $ and $ j+1 $ by $ b_{j,j+1}=(S_j^+S_j^-) (S_{j+1}^+S_{j+1}^-) $. Each $ b_{j,j+1} $ takes $ 0 $ or $ 1 $, but one can easily see that adjacent bond variables $ b_{j-1,j} $ and $ b_{j,j+1} $ can never be $ 1 $ simultaneously. The configuration of $ b_{j,j+1} $ corresponds to the ground state of the model in Ref.~\cite{Lesanovsky2011} by identifying $ b_{j,j+1}=1\leftrightarrow \ket{\uparrow}_j  $ and $ b_{j,j+1}=0\leftrightarrow\ket{\downarrow}_j $. It is an open question whether we can apply similar identification to higher-spin cases.
    
    Our work suggests a number of future research directions. The unperturbed Hamiltonian has infinite number of Onsager-algebra elements commuting with each other. This implies that we could construct other models using such higher Onsager-algebra elements. The generalization to multi-parameter coherent states discussed in \cite{Note2} is one of such examples. Moreover, we could construct a Floquet scar~\cite{Sugiura2019, Mukherjee2019} with Hermitian Onsager-algebra elements.

\medskip
	
\textit{Acknowledgments.---}%
    We thank  Keiji Saito, Eiki Iyoda, Takashi Mori, Naoto Shiraishi, and Ryusuke Hamazaki, for fruitful discussions. We are also thankful for the help of numerical calculation library QuSpin~\cite{Weinberg2017, Weinberg2019} and QuTip~\cite{Johansson2012, Johansson2013}.
    N.S. acknowledges the support by the Materials Education program for the future leaders in Research, Industry, and Technology (MERIT). 
    N.Y. was supported by Advanced Leading Graduate Course for Photon Science (ALPS) of Japan Society for the Promotion of Science (JSPS). N.Y. was supported by JSPS KAKENHI Grant-in-Aid for JSPS fellows Grant No. JP17J00743.
    H.K. was supported in part by JSPS Grant-in-Aid for Scientific Research on Innovative Areas No. JP18H04478 and JSPS KAKENHI Grant No. JP18K03445.

\bibliography{note}

\pagebreak
\widetext
\begin{center}
	\textbf{\large Supplemental Materials for: Onsager's scars in disordered spin chains}
\end{center}
\setcounter{section}{0}
\setcounter{equation}{0}
\setcounter{figure}{0}
\setcounter{table}{0}
\setcounter{page}{1}
\renewcommand{\theequation}{S\arabic{equation}}
\renewcommand{\thefigure}{S\arabic{figure}}

\section{Self-duality and Onsager symmetry of the Hamiltonian (1)}\label{sec:onsager_algebra}
    In this section, we review the self-dual U(1)-invariant clock model in more detail. We start with the original Hamiltonian in Ref.~\cite{Vernier2019}
    \begin{align}
        H_{\mathrm{orig}, n}&=\mathrm{i}\sum_{j=1}^{L}\sum_{a=1}^{n-1}\dfrac{1}{1-\omega^{-a}}\qty[(2a-n)\qty(\tau_j^a+(\sigma_j^\dagger \sigma_{j+1})^a)+\sum_{b=1}^{n-1}\dfrac{1-\omega^{-ab}}{1-\omega^{-b}}\qty(\tau_j^a(\sigma_j^\dagger \sigma_{j+1})^b+(\sigma_j^\dagger \sigma_{j+1})^b\tau_{j+1}^a)]\label{eq:FZ_clock}\\
		&=\mathrm{i}\sum_{j=1}^{L}\sum_{a=1}^{n-1}\dfrac{1}{1-\omega^{-a}}\qty[(2a-n)\tau_j^a+n(S_j^+S_{j+1}^-)^{n-a}-n(S_j^- S_{j+1}^+)^a]\label{eq:FZ_spin}.
    \end{align}
    under periodic boundary conditions (PBC). First, as its name suggests, the model is \textit{self-dual}, i.e., invariant under the duality transformation
	\begin{align}
		\tau_j\to \sigma_j^\dagger \sigma_{j+1},\; \sigma_j^\dagger \sigma_{j+1}\to \tau_{j+1}
	\end{align}
	up to boundary terms. Second, the $ \mathrm{U}(1) $-charge
	\begin{align}
		Q\coloneqq \sum_{j=1}^{L}S_j^z=\sum_{j=1}^{L}\sum_{a=1}^{n-1}\dfrac{1}{1-\omega^{-a}}\tau_j^a
	\end{align}
	commutes with $ H_n $.  Due to these two properties, $ H_n $ also commutes with the dual of $ Q $:
	\begin{align}
		\hat{Q}\coloneqq \sum_{j=1}^{L}\sum_{a=1}^{n-1}\dfrac{1}{1-\omega^{-a}}(\sigma_j^\dagger \sigma_{j+1})^a.
	\end{align}
	Ref.~\cite{Vernier2019} found that $ Q $ and $ \hat{Q} $, which do not commute, generate the \textit{Onsager algebra}. In fact, they satisfy the Dolan-Grady relation~\cite{Dolan1982}
	\begin{align}
	\begin{split}
	    [Q, [Q, [Q, \hat{Q}]]]=n^2[Q, \hat{Q}],\\
	    [\hat{Q}, [\hat{Q},[\hat{Q},Q]]]=n^2[\hat{Q},Q],
	\end{split}
	\end{align}
	which is known as a necessary and sufficient condition to generate the Onsager algebra. For Onsager's original notation, $ A_0=(4/n)Q $ and $ A_1=(4/n)\hat{Q} $ generate $ A_n $ and $ G_n $ which obey
	\begin{align}
	    [A_m, A_n] &= 4G_{m-n},\\
	    [A_m, G_n] &= 2A_{m-n}-2A_{m+n},\\
	    [G_m, G_n] &= 0.
	\end{align}
	Moreover, it follows from $ [Q, H_n]=[\hat{Q}, H_n]=0 $ that $ H_n $ commutes with all the Onsager-algebra elements.
	It may be useful to define $ Q_m^0=(A_m+A_{-m})/2 $ and $ Q_m^\pm=(A_m-A_{-m}\mp 2G_m)/4 $ and remark
	\begin{align}
	    [Q_l^r, Q_m^r] &= 0\quad (r=0, \pm),\\
	    [Q_l^-, Q_m^+] &= Q_{m+l}^0-Q_{m-l}^0,\\
	    [Q_l^-, Q_m^0] &= 2(Q_{m+l}^- -Q_{m-l}^-),\\
	    [Q_l^+, Q_m^0] &= 2(Q_{m-l}^+ -Q_{m+l}^+).
	\end{align}
	$ Q^+ $ in the main text corresponds to
	\begin{align}
	    \frac{n}{4}Q_1^+=\sum_{j=1}^L \sum_{a=1}^{n-1} \frac{1}{1-\omega^a} (S_j^+)^a(S_{j+1}^+)^{n-a}
	\end{align}
    up to constant after unitary transformation explained below.
	
	The Hamiltonian (1) can be obtained by the unitary transformation:
	\begin{align}
	    H_n=U^{-1} H_{\mathrm{orig}, n}U,\quad  U=\exp[\mathrm{i}\pi\qty(1+\dfrac{1}{n})\sum_{j=1}^LjS_j^z].
	\end{align}
	Note that PBC in the original $ H_{\mathrm{orig},n} $ are twisted in $ H_n $. However, even if we impose PBC on $ H_n $, every commutation relation is valid after unitary transformation and redefining $ S_{L+1}^r $ as $ S_1^r $ ($r=z, \pm $). (If $ n $ is even we further assume even $ L $.)

\section{$ \exp(\beta^n Q^+) $ as a matrix product operator}
    Here, we derive the matrix product operator (MPO) representation of $ \exp(\beta^n Q^+) $, which immediately leads to an MPS representation of a coherent state. The key observation is that the series expansion of each local term of $ Q^+ $ in $ \exp(\beta^n Q^+) $ is finite:
    \begin{align}
    \begin{split}
        \exp(\beta^n Q^+)&=\prod_{j=1}^L \qty(1+\beta^n\sum_{a=1}^{n-1}\dfrac{(-1)^{(n+1)j+a}}{\sin(\pi a/n)}(S_j^+)^a (S_{j+1}^+)^{n-a})\\
        &=
        \begin{pmatrix}
            1 & \beta^{n-1}\dfrac{(-1)^{(n+1)+(n-1)}}{\sin[\pi(n-1)/n]}(S_1^+)^{n-1} & \cdots & \beta \dfrac{(-1)^{(n+1)+1}}{\sin(\pi/n)}S_1^+
        \end{pmatrix}
        \begin{pmatrix}
            1\\
            \beta S_2^+\\
            \vdots\\
            (\beta S_2^+)^{n-1}
        \end{pmatrix}\\
        &\hspace{1em}\times
        \begin{pmatrix}
             1 & \beta^{n-1}\dfrac{(-1)^{2(n+1)+(n-1)}}{\sin[\pi(n-1)/n]}(S_2^+)^{n-1} & \cdots & \beta \dfrac{(-1)^{2(n+1)+1}}{\sin(\pi/n)}S_2^+
        \end{pmatrix}
        \begin{pmatrix}
            1\\
            \beta S_3^+\\
            \vdots\\
            (\beta S_3^+)^{n-1}
        \end{pmatrix}\\
         &\hspace{1em}\times\dots\times
          \begin{pmatrix}
             1 & \beta^{n-1}\dfrac{(-1)^{(n+1)L+(n-1)}}{\sin[\pi(n-1)/n]}(S_L^+)^{n-1} & \cdots & \beta \dfrac{(-1)^{(n+1)L+1}}{\sin(\pi/n)}S_L^+
        \end{pmatrix}
        \begin{pmatrix}
            1\\
            \beta S_1^+\\
            \vdots\\
            (\beta S_1^+)^{n-1}
        \end{pmatrix}\\
        &=\tr(C^{[1]}\dots C^{[L]}),
    \end{split}
    \end{align}
    where
    \begin{align}
        (C^{[l]})_{ij}&=
        \begin{dcases}
            (\beta S_l^+)^i & (j=0)\\[2ex]
            \dfrac{(-1)^{(n+1)l+(n-j)}}{\sin[\pi(n-j)/n]}(\beta S_l^+)^{n+i-j} & (j\ne 0)
        \end{dcases}
    \end{align}
    for $ 0\le i,j \le n-1 $ (0-based indexing).

\section{Multi-parameter Coherent State}
    A coherent state in the main text can be generalized to a multi-parameter one. One obtains the following Onsager algebra elements (see also Sec.~\ref{sec:onsager_algebra} in the Supplemental Materials) by straightforward calculation:
    \begin{align}
        Q_{l}^+\propto \sum_{j=1}^{L} (-1)^{j+1}S_{j}^+ \qty(\prod_{k=j+1}^{j+l-1} S_{k}^z)S_{j+l}^+.
    \end{align}
    $ Q_1^+ $ corresponds to $ Q^+ $ in the main text. Then, we can construct a multi-parameter coherent state
    \begin{align}
        \ket{\psi(\beta_1,\dots,\beta_m)}\coloneqq \qty(\prod_{l=1}^m \exp(\beta_l^2 Q_l^+))\ket{\Downarrow}.
    \end{align}
    Note that this state does not have overlap with
    \begin{align}
        \ket{\underbrace{0\dots 0}_{m}1\underbrace{0\dots 0}_{m}}
    \end{align}
    over $ (2m+1) $ consecutive sites. Thus, we can construct a QMBS Hamiltonian by adding a perturbation term $ \sum_{j=1}^L c_j\ket{0\dots 010\dots 0}_{j-m,\dots, j+m}\bra{0\dots 010\dots 0} $. As a more exotic situation than that in the one-parameter case, we have $ \order{N^m} $ scar states written as $ \qty(\prod_{l=1}^m (Q_l^+)^{k_l})\ket{\Downarrow} $, where $ k_l\in \mathbb{N} $.
    
\begin{figure}[H]
        \centering
        \includegraphics[width=0.5\linewidth]{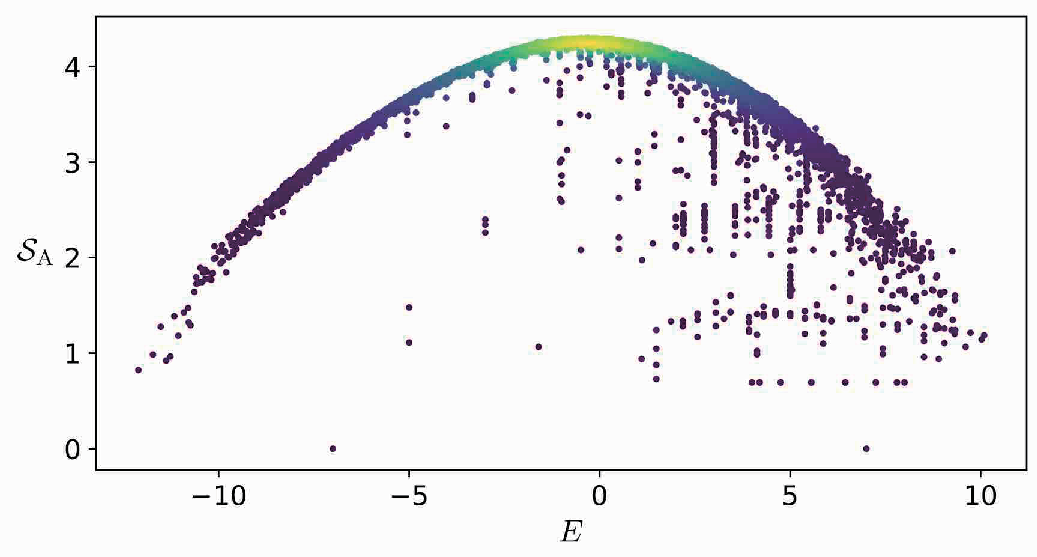}
        \caption{Half-chain bipartite EE for $ n=2,\, L=14 $ with perturbations consisting of states in Eq.~(\ref{eq:pert_higher_Onsager}).}
        \label{fig:higher_Onsager}
    \end{figure}

    Let us demonstrate the above construction in the two-parameter case. Using $ Q_1^+ $ and $ Q_2^+ $, we can construct a two-parameter coherent state
    \begin{align}
        \ket{\psi(\alpha, \beta)}\coloneqq \exp(\alpha^2 Q_1^+)\exp(\beta^2 Q_2^+)\ket{\Downarrow}.
    \end{align}
    Similarly to the one-parameter one, this two-parameter one can be written as an MPS state, since $ \exp(\alpha^2 Q_1^+)\exp(\beta^2 Q_2^+) $ can be written as an MPO:
    \begin{align}
         \exp(\alpha^2 Q_1^+)\exp(\beta^2 Q_2^+)=\tr(D^{[1]}\dots D^{[L]}),
    \end{align}
    where
    \begin{align}
        (D^{[j]})&=
        \begin{dcases}
            \begin{pmatrix}
                1&\alpha S_j^+\\
                \alpha S_j^+&0
            \end{pmatrix}
            \otimes
             \begin{pmatrix}
                1&\beta S_j^+\\
                \beta S_j^+&0
            \end{pmatrix}
            \otimes
             \begin{pmatrix}
                1&0\\
                0&-S_j^z
            \end{pmatrix}
            & (l:\text{odd})\\[2ex]
           \begin{pmatrix}
                1&-\alpha S_j^+\\
                \alpha S_j^+&0
            \end{pmatrix}
            \otimes
             \begin{pmatrix}
                1&0\\
                0&S_j^z
            \end{pmatrix}
             \otimes
             \begin{pmatrix}
                1&\beta S_j^+\\
                \beta S_j^+&0
            \end{pmatrix}
            & (l:\text{even})
        \end{dcases}.
    \end{align}
    The MPS representation tells us that $ \ket{\psi(\alpha, \beta)} $ does not have overlap with 
    \begin{align}
        \ket{00100}\quad \text{nor}\quad \dfrac{\ket{00101}-\ket{10100}}{\sqrt{2}}.\label{eq:pert_higher_Onsager}
    \end{align}
   Then, adding the perturbation made from these states, we can construct a QMBS Hamiltonian in a similar way to the main text. Figure \ref{fig:higher_Onsager} shows the EE of this model. One can see at least $ \order{N^m}$ (here $ m=2 $) anomalously low-entangled states.
    
\section{ETH-Violation in physical observables}
\begin{figure}[H]
    \centering
    \includegraphics[width=0.65\linewidth]{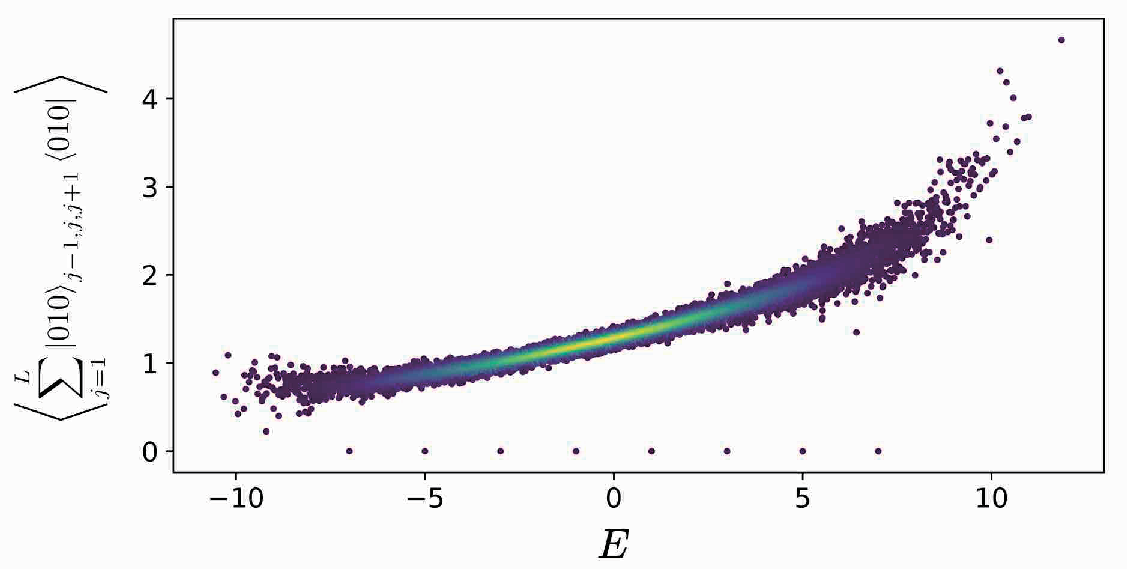}
    \caption{Expectation values of $ \sum_{j=1}^L \ket{010}_{j-1, j,j+1}\bra{010} $ for all energy eigenstates with $ L=14 $ and $ h=1.0 $. Perturbation parameters $ c_j^{(1)},\, c_j^{(2)} $, and $ c_j^{(3)} $ are randomly chosen from $ [-1,1]$. The values for scar states are $ 0 $, and well separated from those for other typical states.}
    \label{fig:expectation_value}
\end{figure}

In this section, we show the plot of expectation values of macroscopic summation of local observables for all energy eigenstates. If the ETH holds, these for each small energy shell are indistinguishable. However, as seen in Fig.~\ref{fig:expectation_value}, it is not the case in our model, which implies the ETH violation.

\section{Results for $ S=1 $ case}
    Here, we show the several results for $ S=1 $ case. The unperturbed Hamiltonian which is similar but different from Ref.~\cite{Chattopadhyay2019} and $ Q^+ $ is
    \begin{align}
        H_3&=\sqrt{3}\sum_{j=1}^L \qty[S_j^-S_{j+1}^+ +S_j^+S_{j+1}^--(S_j^- S_{j+1}^+)^2-(S_j^+S_{j+1}^-)^2-(S_j^z)^2+\dfrac{2}{3}],\\
        Q^+&=\dfrac{2}{\sqrt{3}}\sum_{j=1}^L S_j^+(S_j^+-S_{j+1}^+)S_{j+1}^+.
    \end{align}
    \subsection{Perturbation terms}
\begin{figure}[H]
        \centering
        \includegraphics[width=0.5\linewidth]{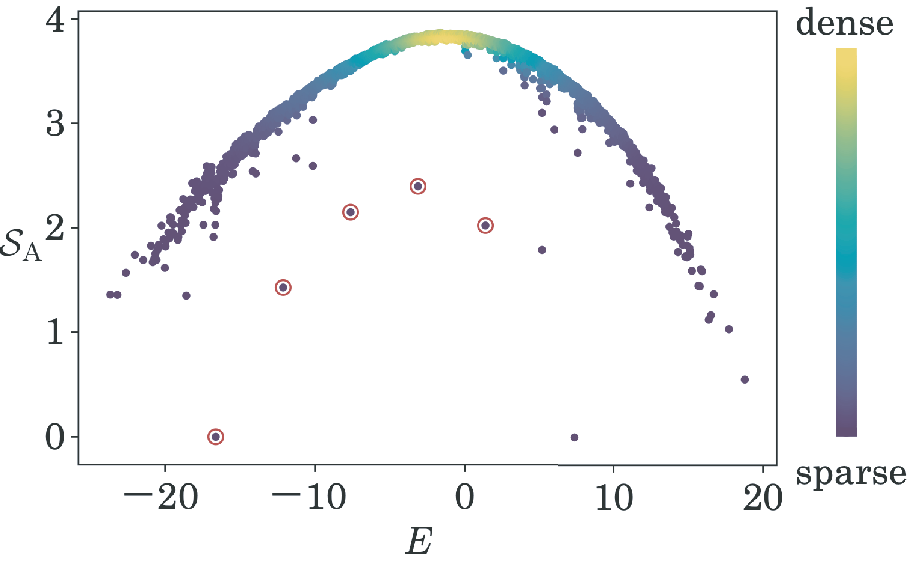}
        \caption{Half-chain bipartite EE for $ n=3,L=8, h=1.5 $ with all perturbations consisting of states in Eq.~(\ref{eq:n3_pert_terms}). The strength of each local perturbation term is chosen randomly. The red circles denote $ (Q^+)^k\ket{\Downarrow} $. One can see that one-magnon states are no longer scar states.}
        \label{fig:EE_n3_wo_one-magnon_scar}
    \end{figure}

    Similarly to the $ S=1/2 $ case, the MPS representation tells us possible perturbations. For $ S=1 $ case, the following states do not have overlap with the coherent state:
    \begin{gather}
        \ket{010},\; \ket{020},\; \ket{110},\; \ket{110},\; \ket{011},\ket{111},\notag\\ \frac{1}{2}\qty(\ket{012}+\ket{021}+\ket{120}+\ket{210}),\; \frac{1}{\sqrt{2}}\qty(\ket{012}-\ket{120}),\; \frac{1}{\sqrt{2}}\qty(\ket{021}-\ket{210}),\notag\\
        \frac{1}{2}\qty(\ket{022}-\ket{112}-\ket{211}+\ket{220}),\; \frac{1}{2}\qty(\ket{022}+\ket{112}-\ket{211}-\ket{220}),\; \frac{1}{2\sqrt{2}}\qty(\ket{022}+\ket{112}+2\ket{121}+\ket{211}+\ket{220}),\notag\\
        \frac{1}{\sqrt{3}}\qty(\ket{122+\ket{212}}+\ket{221})\label{eq:n3_pert_terms}
    \end{gather}
    By using these, we can construct perturbation terms $ H_{\mathrm{pert}, 3} $, which breaks $ \mathrm{U}(1) $ symmetry in general. Note that if we do not use the last one, then perturbation terms become zero for one-magnon states $ \ket{2\dots 212\dots 2} $. That is how we have one-magnon scars in Fig.~2. When $ H_{\mathrm{pert}, 3} $ includes the term using the last one, one-magnon states are no longer scar states as shown in Fig.~\ref{fig:EE_n3_wo_one-magnon_scar}.
    
    \subsection{Dynamics}
        Here, we show numerical results of dynamics of the fidelity and half-chain bipartite EE for the case of $ S=1 $. One can clearly see a similar behavior to the $ S=1/2 $ case, which demonstrates the validity of the construction of our scarred model for the $S=1$ case.
        \begin{figure}[H]
            \centering
            \includegraphics[width=0.5\linewidth]{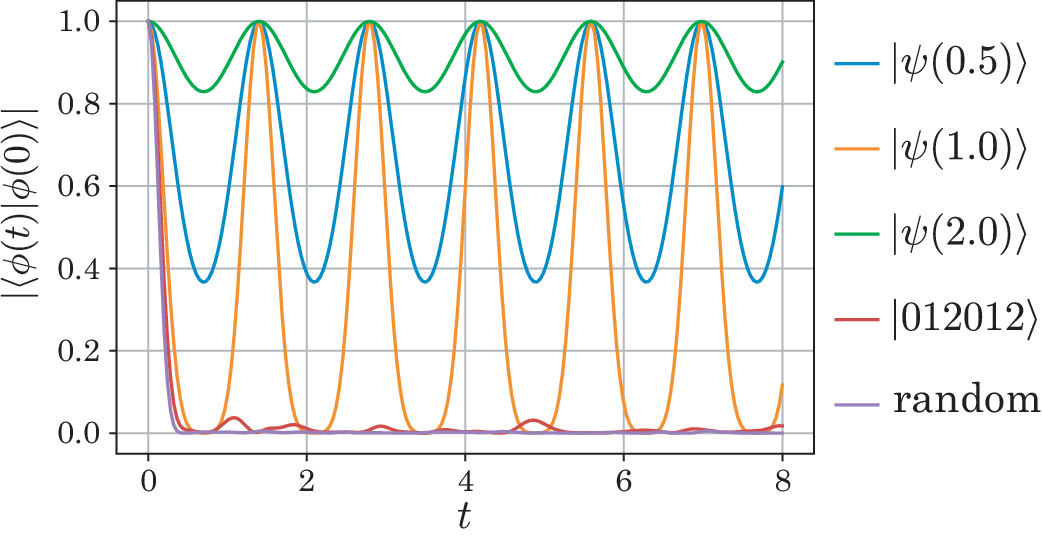}
            \caption{Fidelity dynamics with $ n=3,\; L=6,\; h=1.5 $, and randomly chosen perturbation terms. The coherent states have perfect revivals with period $ 2\pi/(nh)\simeq 1.4 $.}
            \label{fig:fidelity_n3}
        \end{figure}
        \begin{figure}[H]
            \centering
            \includegraphics[width=0.5\linewidth]{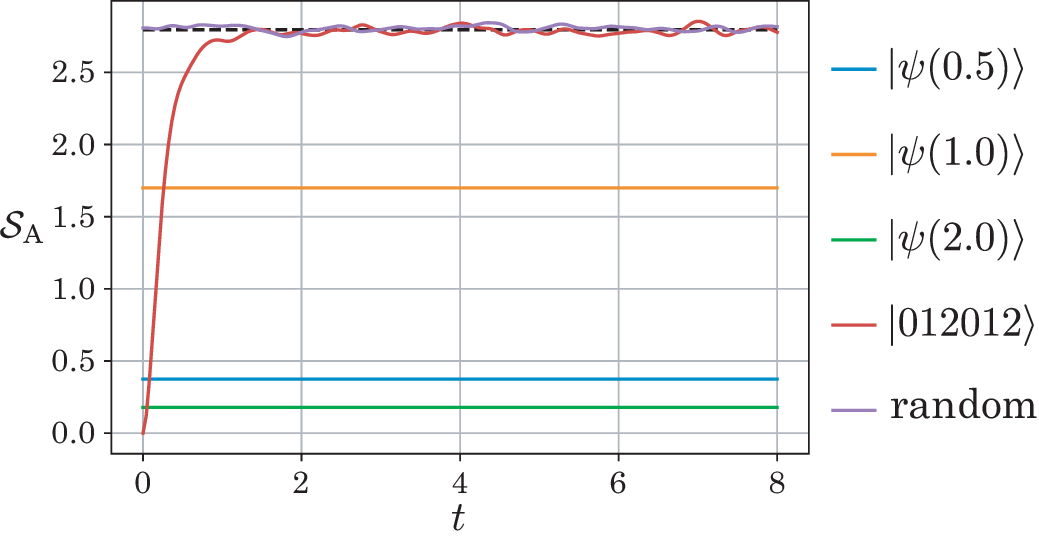}
            \caption{Dynamics of the half-chain bipartite EE with the same setup as Fig.~\ref{fig:fidelity_n3}. The Page value denoted by the black dashed line for $ S=1 $ is $ \mathcal{S}_\mathrm{Page}=(L/2)\ln 3-1/2 $.}
            \label{fig:EE_growth_n3}
        \end{figure}

\section{Sub-volume-law entanglement of $(Q^+)^k\ket{\Downarrow} $}
    Here, we show that the half-chain bipartite EE of $(Q^+)^k\ket{\Downarrow} $ is sub-volume law for $ n=2 $. In this section, we impose the open boundary condition (OBC) for $ Q^+ $. Although our model is valid only under PBC, we expect that the boundary condition does not matter to the scaling of EE. Only in this section, we apply to $ Q^+ $ the unitary transformation to obtain
    \begin{align}
        Q^+\to \sum_{j=1}^{L-1}S_j^+S_{j+1}^+.
    \end{align}
    This unitary transformation is just product of single-site rotations, and hence does not alter the entanglement properties of $ (Q^+)^k\ket{\Downarrow} $.
    
    $(Q^+)^k\ket{\Downarrow} $ can be written as the following MPS:
    \begin{align}
        (Q^+)^k\ket{\Downarrow}\propto\sum_{m1,\dots, m_L}\langle\langle 0| M_{m_1}\cdots M_{m_L}| 2k\rangle\rangle \ket{m_1\dots m_L}\eqqcolon \ket{\xi(k)},
    \end{align}
    where
    \begin{align}
        (M_0)_{ij}=
        \begin{cases}
            \delta_{ij}&i\text{: even}\\
            0&i\text{: odd}
        \end{cases},\quad 
        (M_1)_{ij}=\delta_{i+1, j}
    \end{align}
    is $ (2k+1) \times (2k+1) $ matrices, and $ \langle\langle 0| $ and $ |2k\rangle\rangle $ denote the boundary indices of the auxiliary space. The Schmidt decomposition of $ \ket{\xi(k)}$ is obtained by rewriting it as
    \begin{align}
        \ket{\xi(k)}=\sum_{l=0}^{2k}\qty(\sum_{m_1,\dots,m_{L/2}}\langle\langle 0|M_{m_1}\cdots M_{m_{L/2}}|l\rangle\rangle \ket{m_1\dots m_{L/2}})\otimes \qty(\sum_{m_{L/2+1},\dots,m_L}\langle\langle l| M_{m_{L/2+1}}\cdots M_L |2k\rangle\rangle\ket{m_{L/2+1}\dots m_L}).
    \end{align}
    Then, we calculate the power of the transfer matrix
    \begin{align}
        E_{(a, c)(b,d)}=\sum_{m}(M_m)_{ab}(M_m)^*_{cd}.
    \end{align}
    One can obtain
    \begin{align}
        (E^n)_{(0,0)(l,l)}=
        \begin{cases}
            \max(0, \begin{pmatrix}
                n-l/2\\
                l/2
            \end{pmatrix})&l\text{: even}\\
             \max(0, \begin{pmatrix}
                n-(l+1)/2\\
                (l-1)/2
            \end{pmatrix})&l\text{: odd}
        \end{cases}.
    \end{align}
    In the following, we assume $ L $ is a multiple of $ 4 $. The numerical result suggests that the most entangled state of $ (Q^+)^k\ket{\Downarrow} $ is $ (Q^+)^{L/4}\ket{\Downarrow} $, i.e., a half-filling state, in which case $ \ket{\xi(L/4)} $ can be written as
    \begin{align}
    \begin{split}
        \ket{\xi(L/4)}&=\sum_{l=0}^{L/2}\qty(\sum_{m_1,\dots,m_{L/2}}\langle\langle 0|M_{m_1}\cdots M_{m_{L/2}}|l\rangle\rangle \ket{m_1\dots m_{L/2}})\\
        &\hspace{1em}\otimes \qty(\sum_{m_{L/2+1},\dots,m_L}\langle\langle 0| M_{m_{L/2+1}}\cdots M_L |L/2-l\rangle\rangle\ket{m_{L/2+1}\dots m_L})\\
        &=\sum_{l=0}^{L/4}
        \sqrt{c_{2l} c_{L/2-2l}}\ket{\phi_{\mathrm{L},2l}}\ket{\phi_{\mathrm{R},2l}}+\sum_{l=1}^{L/4}
        \sqrt{c_{2l-1}c_{L/2-2l+1}}\ket{\phi_{\mathrm{L},2l-1}}\ket{\phi_{\mathrm{R},2l-1}}.
    \end{split}
    \end{align}
    Here,
    \begin{align}
        c_l=
        \begin{cases}
            \begin{pmatrix}
            L/2 - l/2\\
            l/2
        \end{pmatrix}&l\text{: even}\\
         \begin{pmatrix}
            L/2 - (l+1)/2\\
            (l-1)/2
        \end{pmatrix}&l\text{: odd}
        \end{cases},
    \end{align}
    and $ \qty{\ket{\phi_{\mathrm{L/R},i}}} $ is an orthonormal set for the left/right-half of the chain. Remarkably, its normalization factor has a simple expression with the help of a generalized Vandermonde identity derived in Ref.~\cite{Gould1956}:
    \begin{align}
        \bra{\xi(L/4)}\ket{\xi(L/4)}=\sum_{l=0}^{L/2}c_lc_{L/2-l}=
        \begin{pmatrix}
            (3/4)L\\
            L/4
        \end{pmatrix}
        \eqqcolon \mathcal{N}.
    \end{align}
    Finally, letting region A be the left half of the chain, we obtain a closed formula for the half-chain bipartite EE $ \mathcal{S}_\mathrm{A} $ of $ (Q^+)^{L/4}\ket{\Downarrow}\propto \ket{\xi(L/4)} $:
    \begin{align}
        \mathcal{S}_\mathrm{A}=-\sum_{l=0}^{L/2}\frac{c_lc_{L/2-l}}{\mathcal{N}}\ln \frac{c_lc_{L/2-l}}{\mathcal{N}}.
    \end{align}
    The key to obtaining an upper bound is the Gibbs' inequality~\cite{Applebaum2002}
    \begin{align}
        -\sum_l p_l \ln p_l\le -\sum_l p_l \ln q_l,
    \end{align}
    which holds for any probability distributions $ \qty{p_l} $ and $ \qty{q_l} $ with equality if and only if $ p_l = q_l $ for all $ l $. By taking $ p_l = c_lc_{L/2-l}/\mathcal{N} $ and $ q_l=1/(L/2+1) $, we obtain
    \begin{align}
        \mathcal{S}_\mathrm{A}\le -\sum_{l=0}^{L/2}\frac{c_lc_{L/2-l}}{\mathcal{N}} \ln(\frac{1}{L/2+1})=\ln(L/2+1)=\order{\ln L}.
    \end{align}

\end{document}